# Formal Quantum Software Engineering

Introducing the Formal Methods of Software Engineering to Quantum Computing

Carmelo R. Cartiere [1]

[1] *Nextsense Srl, Division of Quantitative Physics and Systems Engineering*
  *Via della Rotonda 36, 00186 Rome, RM, Italy*

**Abstract**
Quantum computing (QC) represents the future of computing systems, but the tools for reasoning about the quantum model of computation, in which the laws obeyed are those on the quantum mechanical scale, are still a mix of linear algebra and Dirac notation; two subjects more suitable for physicists, rather than computer scientists and software engineers. On this ground, we believe it is possible to provide a more intuitive approach to thinking and writing about quantum computing systems, in order to simplify the design of quantum algorithms and the development of quantum software. In this paper, we move the first step in such direction, introducing a specification language as the tool to represent the operations of a quantum computer via axiomatic definitions, by adopting the same symbolisms and reasoning principles used by formal methods in software engineering. We name this approach formal quantum software engineering (F-QSE). This work assumes familiarity with the basic principles of quantum mechanics (QM), with the use of Zed (Z) which is a formal language of software engineering (SE), and with the notation and techniques of first-order logic (FOL) and functional programming (FP).

## 1  An Introduction to the Quantum Computing Observable

Observables can be considered the most significant entities of QM because, in a quantum object (QO), each observable completely describes a given attribute (e.g., position, momentum, energy) by encompassing all of its possible states, or eigenstates, in a superposed configuration. In QC systems, QOs have only one observable, the qubit, which superposed configuration is the linear combination of two possible eigenstates. Its quantum state vector, commonly expressed in Dirac's bra-ket notation [5], is, therefore, the linear combination of the two eigenvectors $|0\rangle, |1\rangle$; which associated measurable eigenvalues (the scalars) are 0, 1:

$$\vec{\psi} = |0\rangle + |1\rangle$$

As per its QM counterpart, measuring an observable in a QC system will collapse that observable into one of its eigenstates, that for a qubit are those corresponding to either $|0\rangle$ or $|1\rangle$, with probabilities $c_0, c_1$ [1] [11]:

$$\vec{\psi} = c_0 |0\rangle + c_1 |1\rangle$$

---

[1] The probability for an observable to collapse into any of its states is the squared modulus of the states' corresponding probability amplitudes, which are complex numbers that weight each eigenvector and such that it is $|c_0|^2 + \dots + |c_n|^2 = 1$.

## 1.1 Formalizing the Observable

By the third postulate of QM, an observable that has a finite number of quantum states can be represented via a Hermitian matrix [2]. As such, the three requirements that it must have can be described, with a sound formalism, by adopting strongly typed data and first-order logic [1]; i.e.:

1. it must be a complex square matrix of order n:

$$\mathbb{O}^n : \mathbb{C}^{n \times n}$$

2. it must be equivalent to its own conjugate transpose:

$$\forall c : \mathbb{O}^n \; \exists_1 \, c' : \overline{\mathbb{O}^n}^T \; \bullet \; (c_{ij} = c'_{ij})$$

3. for every eigenvector (or, column) of the matrix the eigenvalue must be a real number; and such that it is the element on the main diagonal of the matrix:

$$\forall V^{n \times 1} : \mathbb{P} \, \mathbb{O}^n \, \exists_1 \, \lambda : \mathbb{R} \; \bullet \; \lambda = c_{jj} \in \mathbb{O}^n$$

In Z, all three requirements can be summarized with the following axiomatic definition satisfying the principle of soundness promoted by FM [3]:

$$\begin{array}{|l} \mathbb{O}^n : \mathbb{C}^{n \times n} \\ \hline \forall c : \mathbb{O}^n \; \exists_1 \, c' : \overline{\mathbb{O}^n}^T \; \bullet \; (c_{ij} = c'_{ij}) \in \mathbb{C} \wedge c_{jj} \in \mathbb{R} \end{array}$$

## 1.2 The Observable Operators

After having introduced the new type $\mathbb{O}^n$, it is now possible to define the observable operators. They are elementary quantum gates that perform unitary transformations $U_f$ (i.e., reversible computations) and that, applied to an observable, make it possible to write quantum programs.

In the following paragraphs, we introduce the axiomatic definition of the most common quantum gates, establishing the basic tools to design quantum programs in Z.

**Identity gate.**
It is the simplest, single qubit, quantum operator, that maps the input to the output unchanged. It is required by any operation where the same qubits that are passed as arguments must be returned:

$$\begin{array}{|l} I : \mathbb{O}^2 \rightarrowtail \mathbb{O}^2 \\ \hline \forall x : \{0,1\} \exists r : \{|x\rangle \mapsto |x\rangle\} \; \bullet \; r \in I \Leftrightarrow \Big(|0\rangle \mapsto |0\rangle \wedge |1\rangle \mapsto |1\rangle\Big) \in r \end{array}$$

**Pauli-X (or, Bit Flip) gate.**
It is the quantum equivalent of the classical NOT gate:

---

[2] But if the Hilbert space $\mathcal{H}$ is infinite-dimensional, the observable is described by a *symmetric operator*, which is represented as a map $f$ between two domains of basis' states $D$ and $D^*$ dense in $\mathcal{H}$, such that $\forall x: D, y: D^* \exists f: D \mapsto D^* \bullet \langle f(x), y \rangle = \langle x, f(y) \rangle$. This is a *bijective function* (injective-surjective), in the sense that it cannot map two distinct states of the domain $D$ onto the same state of the co-domain $D^*$, thus preserving its unitary quality. However, because an infinite-dimensional space is unbounded also the operator is unbounded; therefore, it does not have a largest eigenvalue, leaving us with the conclusion that it might not be defined everywhere and, as such, classifying it as a *partial bijective* function, which implies graph inclusion: $D \leq D^*$.

$$X: \mathbb{O}^2 \rightarrowtail \mathbb{O}^2$$
$$\forall x: \{0,1\} \exists r: \{|x\rangle \mapsto |x'\rangle\} \bullet r \in X \Leftrightarrow \left(|0\rangle \mapsto |1\rangle \wedge |1\rangle \mapsto |0\rangle\right) \in r$$

**Phase Shift gate.**
It represents a family of gates, that rotate the basis state $|1\rangle$ of any arbitrary angle $\phi$:

$$R_\phi: \mathbb{O}^2 \rightarrowtail \mathbb{O}^2$$
$$\forall x: \{0,1\} \exists r: \{|x\rangle \mapsto |x'\rangle\} \bullet r \in R_\phi \Leftrightarrow \left(|0\rangle \mapsto |0\rangle \wedge |1\rangle \mapsto |e^{i\phi}\rangle\right) \in r$$

**Pauli-Z (or, $\pi$ Phase Shift) gate.**
It is a special case of the Phase Shift gate, that rotates the basis state $|1\rangle$ a $\pi$ angle:

$$Z: \mathbb{O}^2 \rightarrowtail \mathbb{O}^2$$
$$\forall x: \{0,1\} \exists r: \{|x\rangle \mapsto |x'\rangle\} \bullet r \in Z \Leftrightarrow \left(|0\rangle \mapsto |0\rangle \wedge |1\rangle \mapsto -|1\rangle\right) \in r$$

**Hadamard gate.**
It is perhaps the most useful quantum operator because it maps any basis state to one qubit with balanced superposition, and vice-versa:

$$H: \mathbb{O}^2 \rightarrowtail \mathbb{O}^2$$
$$\forall x: \{0,+,1,-\} \exists r: \{|x\rangle \mapsto |x'\rangle\} \bullet$$
$$r \in H \Leftrightarrow \left(|0\rangle \mapsto |+\rangle \wedge |+\rangle \mapsto |0\rangle\right) \in r \wedge \left(|1\rangle \mapsto |-\rangle \wedge |-\rangle \mapsto |1\rangle\right) \in r$$

**C-Not gate**
The Controlled Not gate is the most popular 2-qubits operator, because it puts two qubits in a separable state, with a tensor product pairing the first qubit with the result of an addition modulo-2 between both. As such, it is used to entangle two qubits or disentangle the EPR pair:

$$N: \mathbb{O}^4 \rightarrowtail \mathbb{O}^4$$
$$\forall x,y: \{0,1\} \exists r: \{|xy\rangle \mapsto |xy'\rangle\} \bullet r \in N \Leftrightarrow \left(|0y\rangle \mapsto |0y\rangle \wedge |1y\rangle \mapsto |1\overline{y}\rangle\right) \in r$$

Similarly to what happens in any conventional computation, a quantum computation is just a sequence of gates applied in a particular order: each gate takes an input and, after having performed its operation on that input, returns an output. However, in QC, the single use of an operator simultaneously applies to all basis states [9].

## 2 A Practical Example of F-QSE: Programming the Deutsch Algorithm from Specifications

The Deutsch algorithm, foundation model of QC [7, 8], proves if a quantum oracle function, i.e. a black-box that performs a unitary transformation $U_f$ on a qubit, is constant (always maximizing the same state) or balanced (returning each state half of the time). It exploits the quantum entanglement principle [6], and requires the use of two quantum operators: a Hadamard gate, for preparing two qubits in balanced superposition, and a C-Not gate, for entangling the two qubits.

In Dirac notation, it is represented as a ket taking a pair of qubits, prepared from two different basis states ($x$ and $y$), and mapping them to an entangled pair where the second qubit performs as the register storing the state (solution) that will be set on the first qubit by the quantum oracle. The measurement of the first qubit shall, therefore, make it collapse into the state that is held by the second qubit, to which it is entangled:

$$|x, y\rangle \xmapsto{U_f} |x, f(x) \oplus y\rangle$$

With the Z notation, the algorithm can be described through axiomatic definitions; either by writing the constraining predicate from the conventional Dirac representation (which is sound but doesn't add much in a SE perspective):

$$\begin{array}{|l} deutsch : (\mathbb{Z}_2 \to \mathbb{Z}_2) \to \mathbb{Z}_2 \\ \hline \forall f : (\mathbb{Z}_2 \to \mathbb{Z}_2); \; x, y : \{0, 1\} \; \bullet \; deutsch(\,f\,) = |x, f(x) \oplus y\rangle \end{array}$$

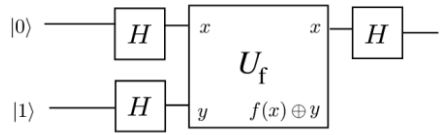

**Figure 1**: The quantum circuit for the Deutsch algorithm.

or from the quantum circuit by taking advantage of the axiomatic definitions of the observable operators, such as:

$$\begin{array}{|l} deutsch : (\mathbb{Z}_2 \to \mathbb{Z}_2) \to \mathbb{Z}_2 \\ \hline \forall f : (\mathbb{Z}_2 \to \mathbb{Z}_2); \; x, y : \{0, 1\} \; \bullet \\ \quad deutsch(\,f\,) = N_f \mid x \leftarrow H \,|0\rangle, y \leftarrow H \,|1\rangle \,\rangle \,\wedge\, H\,|x\rangle \end{array}$$

rather then:

$$\begin{array}{|l} deutsch : (\mathbb{Z}_2 \to \mathbb{Z}_2) \to \mathbb{Z}_2 \\ \hline \forall f : (\mathbb{Z}_2 \to \mathbb{Z}_2); \; x, y : \{0, 1\} \; \bullet \\ \quad deutsch(\,f\,) = x{\leftarrow}H\,|0\rangle \,\wedge\, y{\leftarrow}H\,|1\rangle \,\wedge\, N_f\mid x, y\,\rangle \,\wedge\, H\,|x\rangle \end{array}$$

Indeed, with the last two definitions, by describing the algorithm through a sequence of formal operators we offer a guidance for coding it, by directly following the stepwise logic represented.

Of course, the coding part can be done in any quantum programming language. For our case, in order to match the formal definitions introduced, we worked out an instruction set in Haskell that leans on Green's QIO library [4].

The Deutsch algorithm can be now, effortlessly, translated into the following QC program:

```
deutsch :: (Bool → Bool) → QIO ( Bool )
deutsch f  =  do
x ← qb( "H |0⟩" )
y ← qb( "H |1⟩" )
qN( f ) x y
qH( x )
mq( x )
```

for which, the required operators are implemented as follows:

```
---- return a qubit in a given state
qb :: [Char] -> QIO ( Qbit )
qb qstate
  | qstate = = "|0>" = mkQ( False )
  | qstate = = "|1>" = mkQ( True  )
  | qstate = = "|+>" || qstate = = "H|0>" = do
              qBit <- qb( "|0>" )
              applyU( uhad( qBit ) )
              return( qBit )
  | qstate = = "|->" || qstate = = "H|1>" = do
              qBit <- qb( "|1>" )
              applyU( uhad( qBit ) )
              return( qBit )
  | otherwise = error "qb: wrong argument"

---- apply the C-Not gate to a qubit
qN :: (Bool -> Bool) -> Qbit -> Qbit -> QIO ()
qN f qx qy = applyU( cond (qx) (\ a → if f(a) then unot(qy) else mempty ))

---- apply the Hadamard gate to a qubit
qH :: Qbit -> QIO ()
qH qbit = applyU( uhad( qbit ) )

---- measure a qubit
mq :: Qbit -> QIO ( Bool )
mq qbit = measQ( qbit )
```

## 3   Conclusions and Outlooks

The diffusion of QC cannot be forever relegated within a narrow circle of experts, but many computer scientists and software engineers entering the field of QC are quickly put off by the existing conceptual and notational barriers [2]. This is not only due to the intrinsic difficulty of the subject, but also because it can only be seen through a dark glass (as the complete knowledge of the state of a quantum system is forbidden) [10].

One of the possible ways to overcome this stasis is to introduce a vocabulary inspired by the formal tools of SE. In this paper, the main notions of QC take the form of axiomatic definitions in the Z notation so that they can be used throughout specifications [3]. The result is a notational system that, ideally, can be straightforwardly translated into any quantum programming language.

Hopefully, it can open the doors of QC to a wider audience of players, helping them to beautifully understand, describe, and finally convert the structure of quantum algorithms into fully working code.

## Acknowledgments

The author would like to thank you professor emeritus Prof. Giuseppe Moesch, Prof. Ralf Hinze, and Dr. Wojciech Roga, for the lasting guidance offered during the most intense moments of his life.